\documentclass{article}

\usepackage{arxiv}

\usepackage{amsthm,amssymb,url,amsmath,bm}
\usepackage[font=scriptsize,labelfont=bf]{caption}

\usepackage{float}
\usepackage{graphicx}
\usepackage{color}
\usepackage[utf8]{inputenc} 
\usepackage[T1]{fontenc}
\usepackage{multirow}
\usepackage{tabularx}
\usepackage{textcomp}
\usepackage{relsize}
\usepackage{csquotes}
\usepackage{setspace}
\usepackage{enumerate}
\usepackage{rotating}
\usepackage{bbm}
\usepackage{pdflscape}
\usepackage{todonotes}
\usepackage{cite}
\usepackage[square,sort,comma,numbers]{natbib}
\usepackage{authblk,etoolbox}

\makeatletter
\patchcmd{\maketitle}{center}{flushleft}{}{}                    %
\patchcmd{\maketitle}{center}{flushleft}{}{}                    %
\DeclareMathOperator*{\argmax}{argmax} 
\def\maketitle{{%
  
  \AB@maketitle}}
\makeatother

\usepackage{calc}
\usepackage{tikz}
\usetikzlibrary{decorations.markings}
\tikzstyle{vertex}=[circle, draw, inner sep=0pt, minimum size=6pt]

\title{Emergent inequality and endogenous dynamics in a simple behavioral macroeconomic model}

\author[a,b,c]{Yuki M. Asano}
\author[b,d]{Jakob J. Kolb}
\author[b]{Jobst Heitzig}
\author[e,f,g,1]{J. Doyne Farmer}

\affil[a]{Dept. of Engineering Science, University of Oxford, OX1 3PJ, UK}
\affil[b]{FutureLab on Game Theory and Networks of Interacting Agents, Potsdam Institute for Climate Impact Research (PIK), PO Box 60 12 03, D-14412 Potsdam}
\affil[c]{FernUniversit\"at in Hagen, Universit\"atsstra{\ss}e 47,
D-58097 Hagen}
\affil[d]{Humboldt-Universit\"at zu Berlin, Unter den Linden 6, D-10099 Berlin}
\affil[e]{Institute for New Economic Thinking at the Oxford Martin
  School, University of Oxford, OX1 3UQ, UK} 
  \affil[f]{Mathematical
  Institute, University of Oxford, OX2 6GG, UK}
\affil[g]{Santa Fe Institute, Santa Fe, 87501 NM, USA}
\affil[1]{Corresponding author: doyne.farmer@inet.ox.ac.uk}

\begin{document}
\maketitle

\begin{abstract}
Standard macroeconomic models assume that households are rational in the sense that they are perfect utility maximizers, and explain economic dynamics in terms of shocks that drive the economy away from the stead-state.
Here we build on a standard macroeconomic model in which a single rational representative household makes a savings decision of how much to consume or invest. In our model households are myopic boundedly rational heterogeneous agents embedded in a social network. From time to time each household updates its savings rate by copying the savings rate of its neighbor with the highest consumption.  If the updating time is short, the economy is stuck in a poverty trap, but for longer updating times economic output approaches its optimal value, and we observe a critical transition to an economy with irregular endogenous oscillations in economic output, resembling a business cycle.  In this regime households divide into two groups:  Poor households with low savings rates and rich households with high savings rates.  Thus inequality and economic dynamics both occur spontaneously as a consequence of imperfect household decision making.  Our work here supports an alternative program of research that substitutes utility maximization for behaviorally grounded decision making.
\end{abstract}

conomic growth and inequality are important problems in economics \cite{Acemoglu2009, piketty2015capital}. Standard macroeconomic models are based on the assumption of a single representative rational utility maximizing agent and assume that the dynamics of business cycles are driven by exogenous shocks. However, empirical evidence from behavioral economics indicates that real households are heterogeneous and make substantial deviations from rationality.   This has led to new directions of research, including the incorporation of heterogeneous or boundedly rational agents into macroeconomic models. This is typically done by allowing the agents to differ in terms of factors such as education while preserving the assumption of rationality \cite{ hank_reviewLeahy2018, hank_branch2009new,hank_zhao2018many}, or alternatively allowing for bounded rationality but maintaining utility maximization \cite{hank_gabaix2016behavioral}.  Realism is injected through imposing frictions, such as sticky wages.  These models require shocks to generate economic dynamics.

However, it has long been known that endogenous dynamics are possible in economic models~\cite{day1983emergence,BOLDRIN198626, Scheinkman, boldrin1992equilibrium, %
blume1992evolution}, and more recently Beaudry \textit{et al.}\ have shown how limit cycles can emerge in a standard framework where agents are perfect utility maximizers \cite{beaudry2015reviving, beaudry2016putting}.  An alternative approach bases household decision making on simple heuristics rather than rationality \cite{DeGrauwe2011,DeGrauwe2010the}.  This leads to waves of optimism and pessimism, generating irregular business cycles and giving fat-tailed distributions for economic outcomes such as GDP.   We further develop this line of research by demonstrating how a very simple heterogeneous behavioral macroeconomic model leads to an endogenous business cycle that is not driven by externally imposed shocks.  Our purpose is not to make a fully realistic model, but rather to demonstrate that rich emergent behavior can occur even under very simple assumptions.

Here we extend the Ramsey-Cass-Koopmans (RCK) model, which is one of the foundational models of economic growth theory.  In this model a representative agent rationally chooses a savings rate in order to maximize discounted consumption.  However, there is ample evidence that households do not act as intertemporal optimizing agents and often respond myopically~\cite{Benartzi1995,Loewenstein2000,Choi2016}.  Evidence from lab experiments suggests that individuals perform poorly in finding optimal consumption paths. In reference \cite{carbone2014lifecycle} subjects deviated from optimal consumption choices by roughly 30 percent on average, increasing to roughly 50 percent when subjects were shown the average consumption level in the previous period.  Learning from past generations' consumption paths is somewhat more successful, but the errors are still substantial \cite{ballinger2003precautionary,brown2009learning}.  

We take the opposite approach and assume a strong form of bounded rationality.  In our model households are embedded in a social network and make their savings decisions by simply copying their most successful neighbor. They do this episodically and myopically: From time to time they check all their neighbors and adopt the savings rate of the neighbor with the highest consumption.   

Although we do not claim that this behavior is fully realistic, there is empirical justification for considering a simple rule of this type.  Our agents can be viewed as short-sighted, profligate ``conspicuous consumers'', and the tendency of households to copy one another has been well-documented since the time of Thorstein Veblin \cite{veblen1899}.  Imitate-the-best is one of the decision-making heuristics often applied in settings of high uncertainty and variability~\cite{Gigerenzer2011} and is observed in economic experiments~\cite{traulsen2010human}.  Savings behavior is highly dependent on social interaction with peers~\cite{Lu2011,Zhang2018,Kaustia2012, cascades} and comparing consumption levels incorporates the visibility bias and selection neglect observed in savings rate decisions~\cite{enke2015you}.  Our implementation by copying based on consumption alone is partly motivated by the fact that a neighbor's consumption is more visible than its capital.    This makes it particularly surprising that 
sometimes
their average behavior can be close to optimal.

We find that a key parameter governing economic behavior is the average time interval $\tau$ at which households update their savings rate, which we call the {\it social interaction time}.  When $\tau$ is small, meaning the households update frequently, the savings rate is low, and the performance of the economy is suboptimal in terms of aggregate consumption. When $\tau$ is sufficiently large, in contrast, the economy-wide aggregate savings rate, 
which equals the income-weighted average household savings rate,
becomes close to the optimal rate.  For small $\tau$ the population of households remains homogeneous, but as $\tau$ increases there is a sharp phase transition at a critical value $\tau_{c}$ where the population becomes strongly bimodal, dividing into rich households with high savings rates and poor households with low savings rates.  Correspondingly, for low values of $\tau$ the GDP and other economic indicators are constant with only small fluctuations, whereas above the critical transition there is an endogeneous aperiodic oscillation, resembling a business cycle, in which the aggregate savings rate fluctuates, the population of households alternately becomes richer or poorer, and economic output varies substantially over time.  

Our model shows that the use of heterogeneous agents following explicit behavioral rules can produce aggregate behavior that is qualitatively different from that of rational agents. Our model  is only qualitative, but our results suggest that an approach that explicitly incorporates empirical behavioral knowledge into household decision making may naturally lead to an explanation of business cycles in terms of endogenous dynamics. 

\section*{The standard RCK model}
The RCK model considers a closed economy in which a representative household provides both labor and capital for the production of a single good by a representative firm.  The household receives wages $w$ for labor and a nominal rate of return $r$ on its investment.  It spends a fraction $1\! - \! s$ of its income on consumption and invests the remaining fraction $s$ (see Materials and Methods).

Given the current value of per-capita capital $k$, the household chooses a current value of per-capita consumption $c(s)$ determined by intertemporal optimization, leading to an optimal consumption path that maximizes the household's long-term discounted aggregate utility.  This determines the time evolution of $k$ and $c$ 
towards a steady state at $(k^\ast, c^\ast)$.  

\section*{An agent-based version of RCK}
\subsection*{Economic Model}

We introduce a heterogeneous agent model in the tradition of agent-based modeling \cite{LEBARON19991487,Berry2002,Epstein2006,DOSI20101748, DAWID201454,hommes,Simon2018abmmigration},
using agents that follow a very simple behavioral rule.  
Our model contains $N$ households labeled by $i$ with heterogenous capital $K_i$. For simplicity, all households supply the same labor $L_i = L/N$.  (Introducing heterogeneous labor has little effect on the results).
As in the original RCK model, total economic production is given by the Cobb--Douglas production function, in this case applied to the aggregate input factors $K = \sum_{i=1}^N K_i$ and $L = N L_i$.
As in the original model, capital returns $r$ and wages $w$ equal marginal returns (Methods, Eq.\,\ref{r}),
but incomes $I_i$ now differ between households,
\begin{equation}
\label{Ii}
	I_i = r K_i + w L/N.
\end{equation}

Our key assumption is that each household individually and dynamically sets its time-dependent savings rate $s_i(t)$ according to a behavioral decision rule introduced below,
leading to household capital dynamics
\begin{equation}
\label{kidot}
	\dot{K}_i = s_i I_i - \delta K_i = (r s_i - \delta) K_i + w s_i L/N.
\end{equation}
At the steady state where $\dot{K}_i = 0$, the steady state value $K_i^*$ for household $i$'s capital is a function of the aggregate capital $K$ via its dependence on $w$ and $r$, 
nonlinearly interconnecting all the agents' savings rates and consumption levels.

\subsection*{Household Decision Making}
While the standard RCK model is a one-dimensional dynamical system in which consumption is a deterministic function of the total capital, the agent-based version is $2N$-dimensional, and aggregate consumption depends on all households.
We assume that each household updates its savings rate at random times\footnote{
This leads to smoother transitions than synchronous updates \citep{Vizzari2005, Fates2010}.
}
according to a Poisson process with rate $1/\tau$.  We will see that  $\tau$ plays a crucial role for
the model's behavior. 

Households are embedded in a social network in which each household $i$ has neighbors $\mathcal{N}(i)$.
Whenever household $i$ updates its savings rate, it compares the consumption rates of its neighbors and applies the `imitate-the-best' heuristic,
copying the savings rate of the neighbor with the highest current consumption with a small deviation that can either be interpreted as an error or as an exploration \cite{mehlhorn2015unpacking}.
More precisely,
when the consumption of a neighbor is higher,
it adopts a new savings rate of
\begin{equation}
	s^\mathrm{new}_i = s_{\underset{j\in \mathcal{N}(i)}\argmax_{} (C_j)} + \epsilon,
\end{equation}
where $\epsilon$ is distributed uniformly in the interval of $\pm 1\%$. (The behavior is insensitive to this as long as there is some diversity).

\section*{Results} %

\begin{figure}[tb]
     \centering
       \includegraphics[width=0.7\linewidth]
       {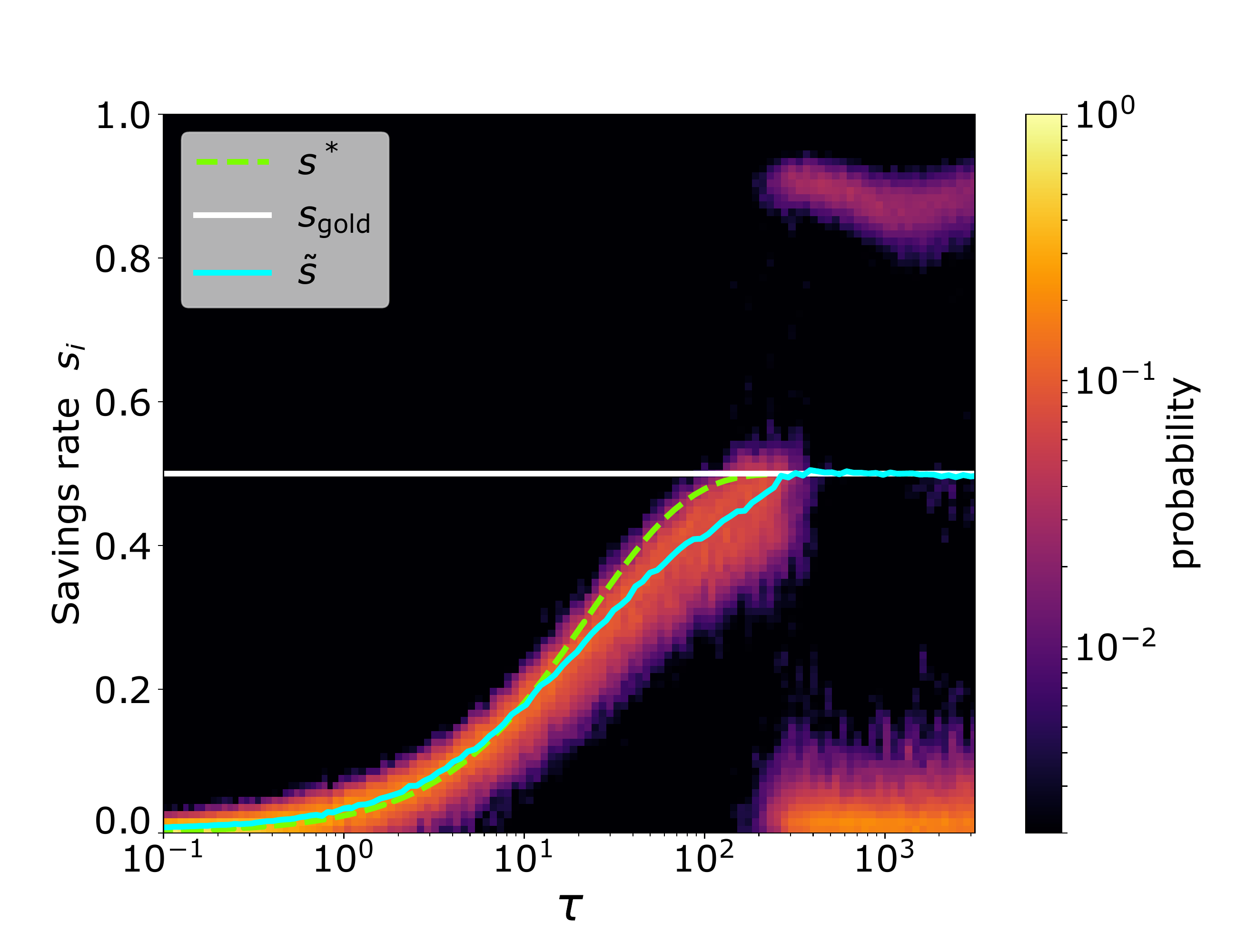}
	\caption []{{ The critical transition from the stable regime to the oscillatory regime.} 
	 We perform an ensemble of simulations at different values of the social interaction time $\tau$, with other parameters held fixed (see SI). We show a heatmap indicating the probability density of the distribution of individual households savings rates for each value of $\tau$, along with the aggregate savings rate $\tilde{s}$. We compare it to the golden rule savings rate $s_\mathrm{gold} \! = 0.5$ and the savings rate $s^*$ predicted by Equation~(\ref{s_optimal}).}

   \label{phase}
\end{figure}

We simulate the model for a variety of different parameters such as the average social interaction time $\tau$ and the network topology. In  Fig.\,\ref{phase} we show the distribution of the final savings rates as a function of the social interaction time $\tau$ for a complete network with the other parameters fixed.  The figure compares this to the optimal, `golden rule' savings rate $s_\mathrm{gold}$,  corresponding to the rational expectations equilibrium where the consumption of the representative agent is maximized.  
There are two distinct regimes, separated by a critical social interaction time $\tau_{c} \approx 250$.
In the {\it stable regime}, corresponding to $\tau < \tau_{c}$, the savings rates of the households are unimodally distributed around a low savings rate.  For very small values of $\tau$ the savings rates are close to zero, and
the economy is stuck in a poverty trap in which its output is very low.  As $\tau$ increases, the savings rate and 
output increase, but the distribution remains unimodal, with a sub-optimal aggregate saving rate.  

For $\tau \! > \! \tau_{c}$ we enter what we call the {\it oscillatory regime}, where the behavior is dramatically different.   In this regime the savings rate distribution is bimodal -- some households have high savings rates and are quite wealthy, while others have low savings rates and are very poor.  We thus observe the spontaneous emergence of extreme inequality, with a lower class and an upper class\footnote{Very near $\tau_\mathrm{c}$ the  distribution in Fig.~\ref{phase} is tri-modal.  This is due to intermittent oscillations between the unimodal and bi-modal regimes.  Thus the system either exhibits a middle class, or a lower class and an upper class, but never all three at once.}.   

Strikingly, as long as $\tau \! > \! \tau_{c}$, the ensemble average of the %
aggregate savings rate $\tilde{s}$ is within $1\%$ of the optimal value $s_\mathrm{gold} \!= \! 0.5$, even when the individual distributions are bimodal.
Furthermore, the time averages of total economic output $Y(t) \!=\! 10.15$ and  consumption $C \! = \! 4.99$ are close to their optimal values $Y^\ast \! = \! s_\mathrm{gold} L/\delta \! = \! 10$  and $C^\ast \! = \! (1-s_\mathrm{gold})Y^\ast \! = \! 5$  in the standard RCK model.  It seems surprising that such a simple, near zero-intelligence learning rule can maintain the system this close to its optimal behavior.

\begin{figure}[htb]
     \centering
       \includegraphics[width=0.7\linewidth]
       {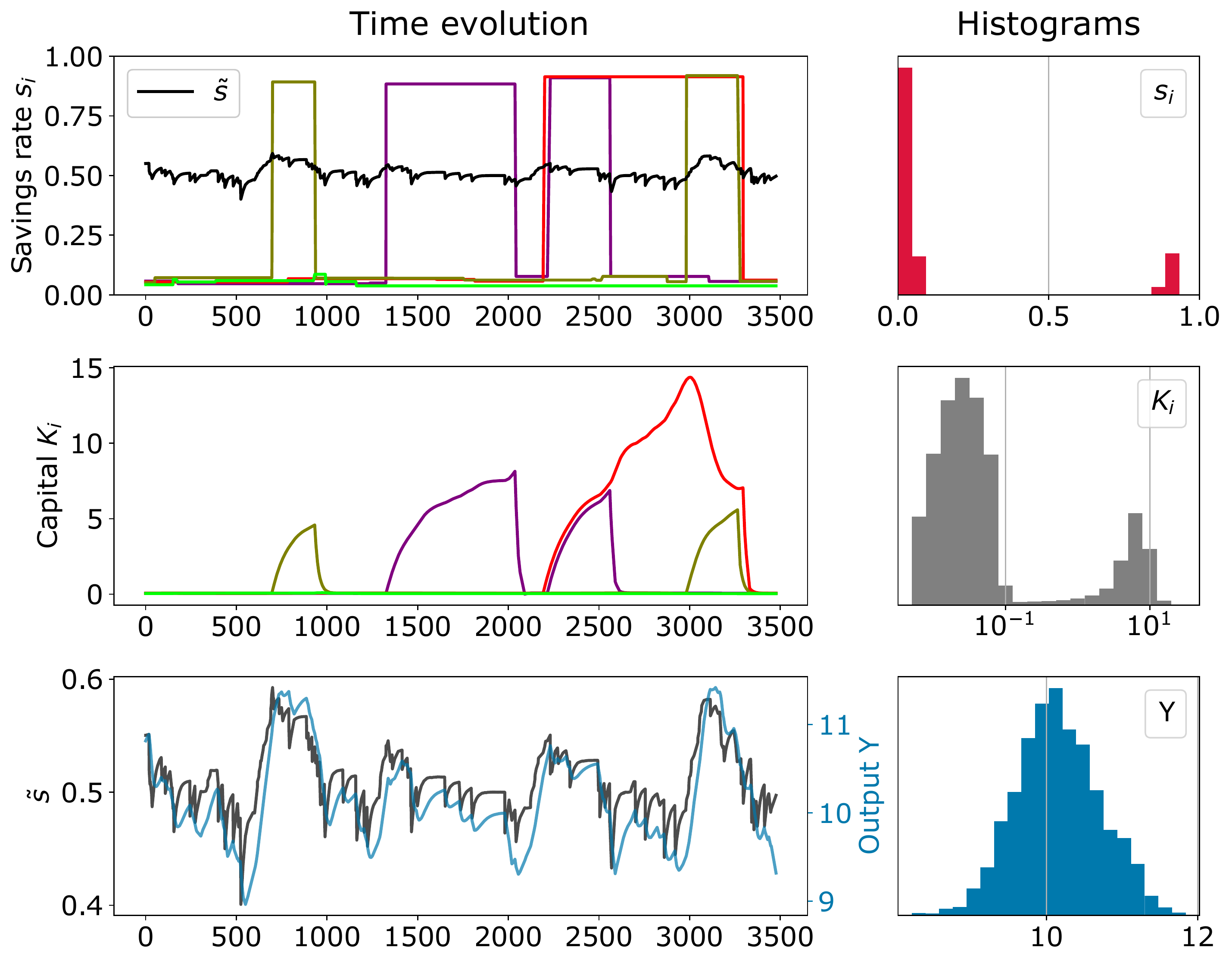}
	\caption{{ The endogenous business cycle in the oscillatory regime}.   We show several time series when $\tau \!>\! \tau_\mathrm{c}$.  
	The top left panel shows the savings rates $s_i(t)$ for four randomly chosen households as a function of time, as well as the aggregate savings rate $\tilde{s}$.  
	The middle left panel shows the capital $K_i(t)$ of the same four households as a function of time. The bottom left panel shows the cyclic behavior of the aggregate output superimposed on the aggregate savings rate.  The panels on the right are histograms of the indicated variables, accumulated over a longer interval.}
   \label{fig:micro_trajs}
\end{figure}

The system dynamics become clearer when we look at the %
economy as a function of time, as illustrated in Fig.~\ref{fig:micro_trajs}.  
For $\tau \! > \! \tau_\mathrm{c}$ there is an endogenous oscillation in many of the aggregate properties of the economy, including the aggregate savings rate $\tilde{s}(t)$ and output $Y(t)$.  This oscillation is also visible in the behavior of individual households.  If we follow any single household it goes through epochs with a high savings rate, near $s_i \! \approx \! 90\%$, and a low savings rate, near $s_i \! \approx \! 5\%$.  At any point in time there is typically an imbalance between rich households and poor households, so that the aggregate savings rate and the aggregate output fluctuate.  We loosely refer to this endogenous oscillation as a ``business cycle''.
\begin{figure*}[h]
     \centering
       \includegraphics[width=0.98\textwidth]
       {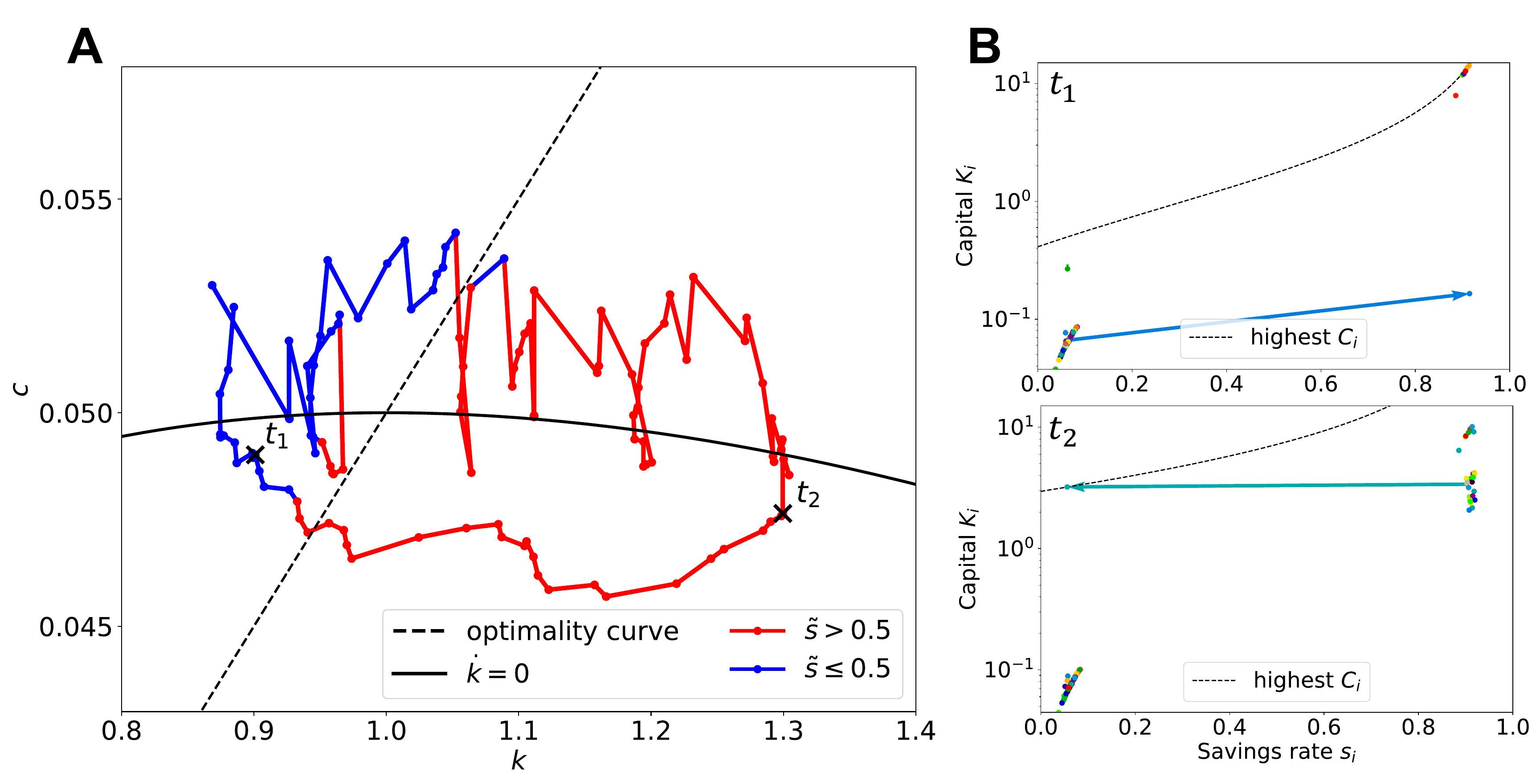}
	\caption{{ Endogenous dynamics in the oscillatory regime. } \textbf{A}:  We plot the average per-capita consumption $c$ against the average per-capita capital $k$ and show the aggregate saving rate $\tilde{s}$ as red when it is greater than $0.5$ and blue when it is less than $0.5$.  The trajectory orbits around the optimal steady state $(k^\ast,c^\ast)$  of the standard RCK model, which is at the intersection of the dashed optimality curve and the solid black $\dot{k}=0$ line. Each dot corresponds to one timestep; the orbit is counterclockwise.
\textbf{B}:~An illustration of the cause of the oscillatory dynamics.  The two panels show snapshots at two different times as indicated in figure A.  At time $t_1$ the aggregate savings rate is low, aggregate capital is low and the economy is in a depression; at time $t_2$ the opposite is true.  The capital and savings rates of individual households are shown as dots with different colors.  There are two clusters, corresponding to rich and poor households.  The household that is currently switching its savings rate is indicated by an arrow connecting its previous state to its current state.  The dashed black curve indicates the iso-consumption curve for the household $i$ with the highest consumption. }
\label{fig:dynamics}
\end{figure*}

\subsection*{Understanding the stable regime ($\bm{\tau \! < \! \tau_{c} }$)}

Although our behavioral rule requires minimal intelligence, the selection process of copying the household with the highest consumption provides a simple mechanism of collective search that becomes more effective as the social updating time $\tau$ increases.  This is perhaps counter-intuitive, as it means that inattention results in superior collective outcomes.  The underlying explanation is as follows:  The  savings rate of the household that is copied has on average been fixed for a time interval of order $\tau$.   When $\tau$ is small, planning is too myopic, ``short term thinking'' dominates, and the households cannot escape using low savings rates with high consumption.  As $\tau$ gets bigger, however, the time between updates becomes long enough that there is more time to accrue an advantage by saving, which drives the savings rate up and increases economic output.  The competitive selection process guarantees that for a sufficiently large population and large $\tau$ the savings rates are close to optimal.%

We use this intuition to derive an approximate formula for the aggregate savings rate $s^*$ as a function of~$\tau$.  We take advantage of the fact that in the stable regime the distribution is unimodal and assume that all households have essentially the same savings rate, and derive the optimal savings rate for time horizon $\tau$.  As explained in detail in the Supplementary Information, for capital elasticity $\alpha=0.5$ the optimal savings rate under these conditions is
\begin{equation}
\label{s_optimal}
s^\ast(\tau) = \frac{1 - e^{-\delta \tau/2}}{2 - e^{-\delta \tau/2}}.
\end{equation}
This approximation is shown in green in Fig.~\ref{phase} and provides a good fit 
throughout the stable regime.

The optimal savings rate in the classical RCK model depends on the discount rate $\rho$, which is a free parameter.  As shown in the SI, substituting $s^\ast$ from Eq.~[\ref{s_optimal}] into the relation for the classical RCK model gives an effective discounting rate for our model in terms of the social interaction time $\tau$ and the depreciation rate $\delta$,  
\begin{equation}
   \rho(\tau) = \frac{\delta/2}{e^{\delta \tau/2} - 1}. \label{eq:rhotau}
\end{equation}
In the limit as $\tau \to 0$, the discount rate $\rho \to \infty$, consistent with the observed collectively myopic behavior. But for $\tau \to \infty$, $\rho \to 0$.
Thus in this case the individually myopic households act collectively ``as if'' they were farsighted, with an emergent effective discounting rate $\rho(\tau)$ which is not a free parameter but is rather a function of the social interaction time $\tau$.

\subsection*{Understanding the oscillatory regime ($\bm{\tau \!>\! \tau_{c} }$)}

To get a deeper understanding of what is happening in the oscillatory regime, where $\tau > \tau_{c}$, in Fig.\,\ref{fig:dynamics} we illustrate the collective and individual dynamics.   In Fig.\,\ref{fig:dynamics}A we show the average per capita consumption rate $c$ as a function of the average capital $k$. This illustrates how the aggregate consumption and capital orbit around the optimal steady state $(k^\ast,c^\ast)$  of the standard RCK model, generating a business cycle.  In relation to the optimal savings rate $s^\ast \!=\!0.5$ of the RCK model, the effective aggregate savings rate $\tilde{s}$ is typically greater than $s^\ast$ when the system is below the optimality curve and less than $ s^\ast$ when it is above the optimality curve. This is interesting as the optimality curve is obtained via optimizing household consumption for an infinite horizon, whereas our model has no explicit optimization.  

To understand what is going on at the individual level, in Fig.\,\ref{fig:dynamics}B we plot a snapshot of the capital vs. the savings rate for all households at two different times, $t_1$ and $t_2$.   At time $t_1$ the economy is just beginning to recover from a recession.  There are two clusters of households, corresponding to rich households in the upper right corner and poor households in the lower left corner.  More households are poor, and because the return $r$ is inversely proportional to total capital according to $r \propto K^{-1 + \alpha}$, where here $\alpha = 0.5$, this means that returns to investment are high.  When the household shown in blue gets its chance to update its savings rate, it copies the higher savings rate of one of the rich households, transitions to the right as indicated by the arrow, and begins accumulating capital by saving more.  Other households follow, and eventually the economy reaches the state shown in the lower panel at time $t_2$, where many houses have high savings rates and are rich.   The resulting excess capital makes the returns on savings low, which when combined with their high savings rates, drives the consumption of these households down.   As a result, when one of the rich households gets its turn to update, it copies a household with a low savings rate and goes on a spending spree.  At this point its consumption rate becomes very high, and all of its neighbors copy it, creating a boom in consumption while decreasing the aggregate savings rate.  A majority of households eventually become impoverished and the cycle repeats itself.   
These dynamics are also given as an animation in the Supplementary Material.

\subsubsection*{Critical social interaction time}

What determines the critical social interaction time $\tau_\mathrm{c}$?  The approximation that the imitate-the-best heuristic results in behavior that is optimal over a time horizon $\tau$ helps understand the instability driving the transition. Suppose that an external shock of size $\Delta$ perturbs the aggregate savings rate $\tilde{s}$ away from $s^\ast$, and suppose that household $i$ is allowed to optimize its savings rate $s_i$ while the others hold theirs constant.  A numerical investigation shows that when $\tau \ll \tau_{c}$ the optimal savings rate $s_i$ computed remains close to $\tilde{s}$.  In contrast, when  $\tau \gg \tau_{c}$, if $\Delta > 0$ then the optimal savings rate is very small, with $s_i$ approaching $0$, and if $\Delta < 0$ the optimal savings rate is large, with $s_i$ approaching $1$ (see Fig.~S3).  This happens because when $\Delta > 0$ the aggregate savings rate is high, so the returns on investment are low, which discourages saving.  Similarly, when $\Delta < 0$ the aggregate savings rate is low, so returns on investment are high, which encourages saving.  This destabilizes the unimodel solution around $s^\ast$.  The transition occurs sharply at a parameter value near $\tau_{c}$, though the precise value depends on $\Delta$.

\subsubsection*{Network size and structure}  
\begin{figure}[!htb]
     \centering
       \includegraphics[width=0.35\textwidth]
       {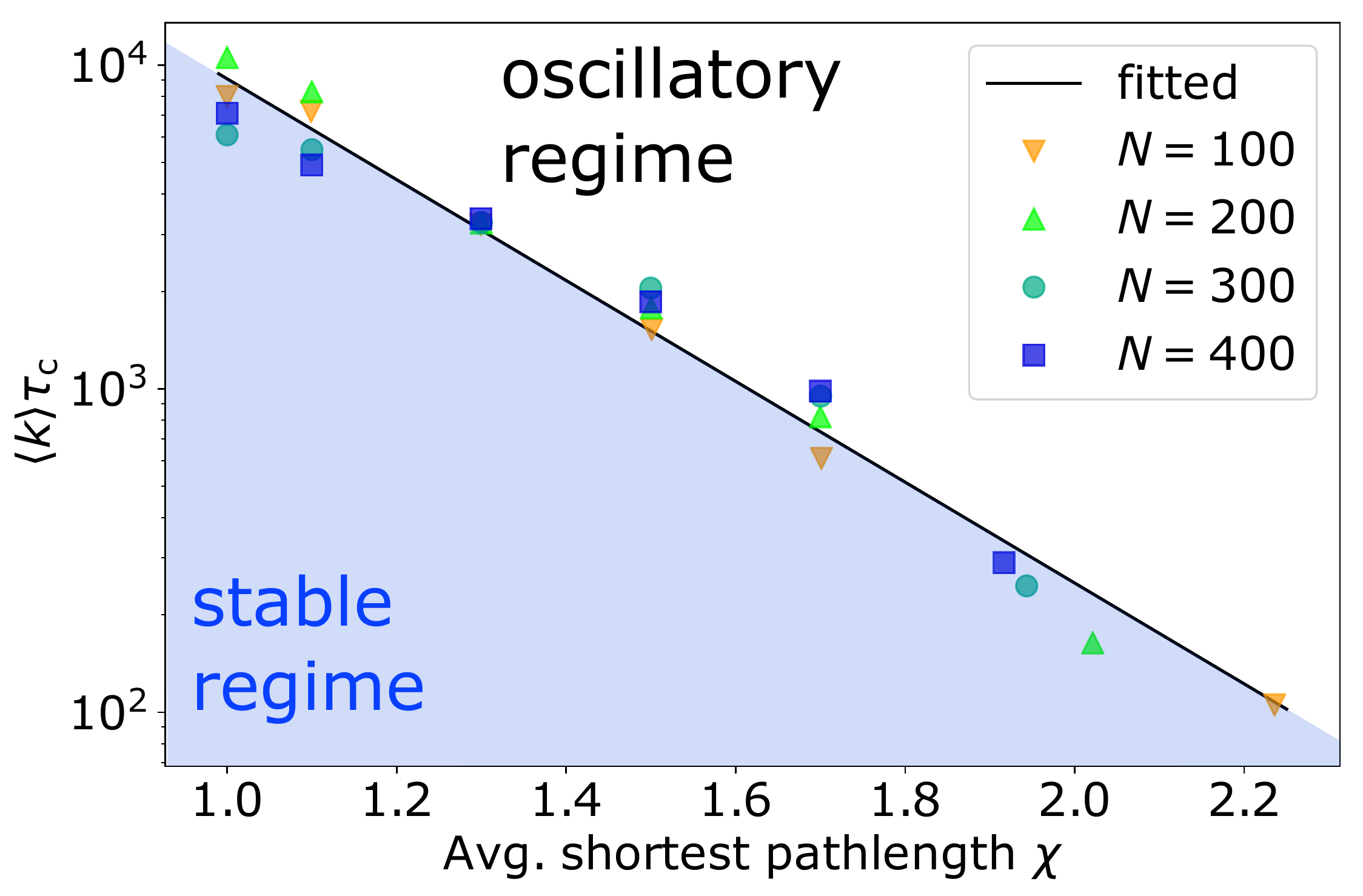}
	\caption{{The critical social interaction time depends on network properties.}
	The logarithm of the mean number of neighbors $\langle k \rangle$  times the critical interaction time $\tau_{c}$ is plotted vs. the average shortest path length $\chi$ for various values of $N$ and $p$, confirming Eq.~[\ref{scaling_relation}]. The stable regime ($\tau \! < \! \tau_{c}$) is shaded in blue. 
   \label{taucrit}}
\end{figure}

We have so far used complete networks in the simulations, but in general the behavior depends on the network size and structure.  For example, we investigate Erdős–Rényi networks with average degree $\langle k \rangle = Np$, where $p$ is the probability that any two nodes are connected.  The critical social interaction time $\tau_{c}$ depends on both $\langle k \rangle$ and $N$ (see Figs.~S4 and S5).  Starting at any given node and moving one link at a time, the number of neighbors that are reached grows exponentially with time at rate $\langle k \rangle$.  The typical distance required for a disturbance to propagate across the network is the average shortest path length $\chi$, defined as the average number of nodes that must be traversed in order to go from any given node to any other node. Motivated by this logic, 
we investigate the empirical relationship between $\tau_c$, $\chi$, and $\langle k \rangle$, finding the proportionality
\begin{equation}
\tau_c \sim e^{-\chi} / \langle k \rangle.
\label{scaling_relation}
\end{equation}  
Fig.\,\ref{taucrit} shows that this makes a good prediction of $\tau_{c}$.  Because $\chi$ increases with $N$, in the large $N$ limit the system is always in the oscillatory regime. Varying the network size and structure parameters also results in qualitative changes in the nature of the oscillation, affecting its frequency, amplitude and variability.  (See a few examples in the SI).

\section*{Discussion}

Our primary purpose here is to make a conceptual point by demonstrating how emergent inequality and endogenous dynamics can naturally emerge from a heterogeneous behavioral model.  Nonetheless, our model makes the prediction that during recessions savings rates increase before output rises (see Fig.~\ref{fig:micro_trajs}).  
This has been %
observed for private savings in 19 OECD countries \cite{adema2015business}.

Although our model has two random inputs, they are small and very different in character from the shocks that drive the dynamics of standard models. The first random input determines the time at which individual households update their savings rates under the Poisson process.  This must be random to ensure that the order in which households update their savings rates varies.  (A fixed order leads to a static economy).  The second random input is the copying error for the savings rate.  This is small ($1\%$) and its value makes little difference to the behavior.  In contrast to standard shocks, which affect the economy as a whole, both of these inputs are at the level of individual households, and affect each household differently.  For a large number of households the copying errors cancel out but the endogenous dynamics nonetheless persist.  Thus while random inputs are necessary in our model, they do not directly drive booms and recessions as the shocks of standard models do.  This is why we say that the economic dynamics in our model are endogenous.

\begin{figure}[htb]
\centering
   \includegraphics[width=0.07\textwidth]
       {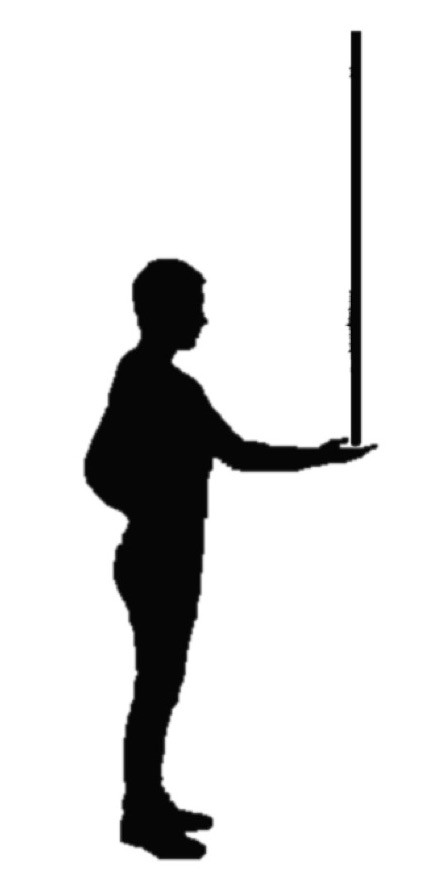}
	\caption[]{{ The problem of pole balancing is analogous to the problem of optimizing savings in an otherwise unstable economy.}  A man attempts to maintain a pole in a vertical position.  This is possible if the pole is long enough, but small errors in the control process drive endogenous oscillations in the angle of the pole.}
   \label{pole_balancing}
\end{figure}

To illustrate the conceptual difference between our model and standard macroeconomic models it is useful to draw an analogy to a simple physical system.  Consider the problem of pole balancing, in which a man attempts to move his hand to maintain a pole in a vertical position, as shown in Fig.\,\ref{pole_balancing}. 
Short poles tip over more quickly than long poles, making it impossible to maintain a vertical position because the pole will tip over before the man can react.  If the pole is long enough, however, the man can move his hand to compensate, and maintain the pole in a roughly vertical position \cite{insperger2017stick}.  There is a sharp critical transition between stability and instability that occurs when the pole is about a meter in length\footnote{This is trivial to confirm empirically -- simply attempt to balance a pole of 60 cm. vs. 130 cm.}.
Nonetheless, even when the pole is very long, it is not possible to maintain a perfectly vertical position, and the 
pole oscillates substantially.

An argument in the style of a standard macroeconomic model would posit that the man is a perfect pole balancer, and any deviations in the angle must be driven by external shocks, such as sharp gusts of wind, that suddenly cause the pole to deviate from vertical.  Under this view, after each shock the man moves his hand perfectly to make the pole vertical again as fast as possible, but before he can achieve this, another shock strikes it, making the pole oscillate around its vertical position.  For pole balancing it is clear that this explanation is wrong. Instead, theories that assume that oscillations are endogenously caused by imperfect control provide a better explanation \cite{insperger2017stick}.

Our suggestion here is that we should revisit the conceptual explanation for business cycles as well.  In the analogy above, the position of the man's hand is like the collection of household savings decisions, and the pole/gravity system is like the economy.  Our model adds weight to the idea that at least part of the variation in savings and investment that occurs during business cycles emerges endogenously due to the imperfect reasoning of households and firms.  Our model also suggests that models with heterogeneity might help illuminate the interaction between business cycles and inequality.  The fact that such rich behavior emerges from such a simple model supports a research agenda for macroeconomics based on empirically derived behavioral rules.

\section*{Materials and Methods}
\subsection*{Code}
The source code (Python) will be made available here upon publication.
\subsection*{The standard RCK model}
In our formulation of the standard RCK model, we follow \citep[p.287--317]{Acemoglu2009}, \citep[p.85--135]{Barro2004}  %
and use continuous time, as in the original \citep{Ramsey1928}. We ignore labor growth for brevity.
Using a fixed amount of labor $L$ and the varying amount of capital $K$, the economy produces a single numeraire good $Y$, assuming a Cobb--Douglas production function $Y \! =\! K^\alpha L^{1-\alpha}$ with capital and labor elasticities $\alpha, 1 \!- \! \alpha \in (0,1)$. 
Per-capita production, $y \! =\! Y/L$, is thus a function of per-capita capital, $k \! =\! K/L$, only, $y \! =\! k^\alpha$.
The model assumes fully competitive factor markets and thus the two factors are compensated according to their marginal products, giving wages %
and capital rents %
\begin{equation}
\label{r}
w = {\partial_L Y} = (1 - \alpha) y, 
\quad r = {\partial_K Y} = \alpha y /k ,
\end{equation}
thereby fully redistributing the numeraire good to households and   leaving the representative firm  with no profits.
The main model parameters of interest are the savings rate $s \le 1$ %
and capital depreciation rate $\delta > 0$ that govern aggregate and per-capita capital growth, 
\begin{equation}
\dot{K} = s(rK + wL) - \delta K,
\quad \dot k = \bar{r}k + w - c,
\end{equation}
where $\bar{r}=r-\delta$ is the real return rate and $c=(1-s)(rk+w)$ is per-capita consumption.
The household aims at maximizing its discounted aggregate utility
$\int_0^{\infty} \! \mathrm{d}t \, e^{-\rho t} u(c(t))$,
by choosing an optimal path $s(t)$ for the savings rate, 
where $\rho>0$ is its discount rate. 
For the instantaneous utility, one assumes a constant relative risk aversion (CRRA) function parameterized by $\theta \geq0, \theta \neq 1$, $u(c) = (c^{1-\theta} -1)/(1-\theta)$.
The solution to this problem fulfills the Ramsey--Keynes Equation
that gives the relative consumption growth rate as
$\dot{c}/c = (\bar{r} - \rho) / \theta$,
In particular, this system has two steady states with $\dot c\!=\!0$, a trivial one in which $c\!=\!k\!=\!0$ and another in which 
the real return rate equals the discount rate, $\bar r \!=\! \rho$, corresponding to a modified `golden rule' \citep[p.300]{Acemoglu2009}, with
capital, consumption and savings rate given~by
\begin{equation}
	k^\ast = \left(\frac{\alpha}{\rho + \delta}\right)^{\frac{1}{1-\alpha}}, 
 	\quad c^\ast = {k^\ast}^{\alpha} -  \delta k^\ast,
 	\quad s^\ast_\mathrm{RCK} = \frac{\alpha \delta}{\rho + \delta}.
\end{equation}
For the limit case $\rho\to 0$, this reproduces the Solow model's golden rule \citep[p.35]{Barro2004}, $s^\ast_\mathrm{RCK} \! =s_\mathrm{gold} \!=\! \alpha$, leading to the largest possible sustainable consumption,
$c^\ast \! =\!(1\! -\! \alpha)(\alpha/\delta)^{\alpha/(1-\alpha)}$. %
For $\rho \! > \!0$, the discount rate pushes the households to save less and shift consumption towards the present.
\subsection*{Acknowledgments}
We thank Paul Beaudry, Jean-Philippe Bouchaud, Roger Farmer, Cars Hommes, Ulrike Kornek, Bastian Ott, Martin Braml, Marco Pangallo and Frederik Schaff for valuable comments and suggestions.
YMA is funded by the EPSRC CDT in AIMS (EP/L015897/1) and JDF by Baillie Gifford and the Institute for New Economic Thinking.

\vfill

\pagebreak

\begin{center}
  \textbf{ Supplementary Material: \\ Emergent inequality and endogenous dynamics in a simple behavioral macroeconomic model}\\[.2cm]
  Dated: \today\\[2cm]
\end{center}

\setcounter{equation}{0}
\setcounter{figure}{0}
\setcounter{table}{0}
\setcounter{page}{1}
\renewcommand{\theequation}{S\arabic{equation}}
\renewcommand{\thefigure}{S\arabic{figure}}
\renewcommand{\bibnumfmt}[1]{[S#1]}
\renewcommand{\citenumfont}[1]{S#1}

\section*{Analytical approximation of the stable regime's steady state}

Here we derive a simple approximate formula for the mean overall savings rate that emerges as a stochastic steady state in the stable regime where $\tau < \tau_{\text{c.}}$.

The main idea is to study which member of an ensemble of households starting at similar but slightly different savings rates and capital stocks will have the largest consumption after the short time interval $\tau$, 
and then assume all households will copy the savings rate of this best household with some error. 
This is only an approximation since in the actual model, households do not simultaneously imitate and not after exactly time $\tau$, and the approximation will only be good when households have already converged to similar savings rates and capital stocks.
However, it turns out that it describes rather well the joint motion towards a steady state once households have converged towards each other.
In particular, the steady state savings rate predicted by the approximation can be seen to match the one observed in the numerical observations quite well.

To see how individual households' consumption at time $\tau$ depends on their individual savings rate, we need to approximate the evolution of $r$ and thus of total capital $K$ first.
Assume all households' savings rates $s_i$ are close to the overall savings rate $s$ and stay constant between time zero and time $\tau$. 
Then $K$ evolves as 
\begin{equation}
    \begin{split}
    \dot K &= (r s - \delta) K + w s L \nonumber \\
    	   &= \Big(\alpha\Big(\frac{L}{K}\Big)^{1-\alpha} s - \delta\Big) K + \alpha \Big(\frac{K}{L}\Big)^{1-\alpha} s L.  
	\end{split}
\end{equation}
For $\alpha = 1/2$ this simplifies to 
\begin{equation}
    \label{aggKdotnew}
    \dot{K} = s\sqrt{L K} - \delta K.
\end{equation}
Assuming that $s$ does not change before time $\tau$, this has two solutions given by 
\begin{equation}
\begin{split}
    K(t) &= \Big(\frac{B - E e^{-\delta t/2}}{\delta} \Big)^2, \\
    r(t) &= \sqrt{L/K} / 2 = \frac{A}{B - E e^{-\delta t/2}}, \\
    w(t) &= \sqrt{K/L} / 2 = \frac{B - E e^{-\delta t/2}}{4 A}
\end{split}
\end{equation}
for all $t < \tau$,
where 
$A = \delta \sqrt{L} / 2$,
$B = s\sqrt{L}$,
and
$E$ has the two possible values $s\sqrt{L} \pm \delta \sqrt{K_0}$.
Since we are interested in the case where $r$ is positive, we have $E = s\sqrt{L} - \delta \sqrt{K_0} < B$.

Knowing $r(t)$ and $w(t)$, we can now determine which household consumes most after time $\tau$.
Household $i$'s capital $K_i(t)$ evolves as 
\begin{equation}
\begin{split}
    \dot K_i 
    &= (s_i r(t) - \delta) K_i + w s_i L_i \nonumber \\
    &= \left(\frac{s_i A}{B - E e^{-\delta t/2}} - \delta\right) K_i
        + \frac{B - E e^{-\delta t/2}}{4 A} s_i L_i.
\end{split}
\end{equation}
This has an analytical solution involving complicated hypergeometric functions.
For small values of $\tau$, we can simplify the problem by approximating $r(t)$ and $w(t)$ for $t\in[0,\tau]$ by their mid-term values $r(\tau/2)$ and $w(\tau/2)$, giving
\begin{equation*}
\dot{K_i} \approx G_i K_i + F_i
\end{equation*}
with $G_i = \frac{s_i A}{B - E e^{-\delta \tau/4}} - \delta$
and $F_i = \frac{B - E e^{-\delta \tau/4}}{4 A} s_i L_i$,
which solves as
\begin{equation*}
    K_i(t) \approx (K_i(0) + F_i/G_i)e^{G_i t} - F_i/G_i.
\end{equation*}
The corresponding consumption of household $i$ at time $\tau$ is then
\begin{equation}\label{eq:Citau}
\begin{split}
    C_i(\tau) 
    &= (1 - s_i)(r(\tau) K_i(\tau) + w(\tau) L_i) \nonumber \\
    &\approx (1 - s_i)\left( 
        H((K_i(0) + F_i/G_i)e^{G_i\tau} - F_i/G_i)
        + L_i/4 H
    \right)
\end{split}
\end{equation}
with $H = \frac{A}{B - E e^{-\delta\tau/2}} = \frac{A}{s\sqrt L (1 - e^{-\delta\tau/2}) + \delta \sqrt{K_0} e^{-\delta\tau/2}}$.
Since we assume all households imitate at time $\tau$ that $s_i$ which has led to the largest $C_i(\tau)$,
we can determine whether $s$ will increase or decrease by identifying whether the $s_i$ that gets copied is larger or smaller than $s$.
Since we also assume households' savings rates $s_i$ are distributed closely around $s$ and all $K_i(0), L_i$ are similar, this question can be answered by seeing whether $C_i(\tau)$ increases or decreases when $s_i$ is increased from below $s$ to above $s$, i.e., by studying the derivative $\partial C_i(\tau)/\partial s_i$ at the point $s_i = s$. Up to a factor of $N$, this derivative is
\begin{equation}
     (1 - s) H\left[
        \frac{e^{G\tau} - 1}{G}(L / 4 H - H F/G) + e^{G\tau}\tau H (K_0 + F/G)
    \right] 
    - H[(e^{G\tau} - 1) F/G + e^{G\tau} K_0] 
    - L / 4 H
    \label{sdotnew}
\end{equation}
where $F = \frac{B - E e^{-\delta \tau/4}}{4 A} s L$ and
$G = \frac{s A}{B - E e^{-\delta \tau/4}} - \delta$.
As long as the above expression is positive or negative, $s$ will increase or decrease over time, respectively.

A steady state will then be reached when both $s$ and $K$ change no longer, i.e., when both $\dot K$ as given by Eq.\,[\ref{aggKdotnew}] (with $K=K_0$) as well as $\partial C_i(\tau)/\partial s_i$ as given by Eq.\,[\ref{sdotnew}]
are zero.
The solution of $\dot K = 0$ is 
$K_0 = L s^2 / \delta^2$, 
at which point we have
$E = 0$,
$H = \delta / 2 s$, 
$G = - \delta / 2$, 
$F/G = - L s^2 / \delta^2 = - K_0$, 
$H K_0 = L s / 2 \delta$, and
$H F / G = - L s / 2 \delta = - H K_0$.
Substituting all this into Eq.\,[\ref{sdotnew}] and setting it zero gives the following surprisingly simple approximate equation for the steady state $s$:
\begin{equation}
    s^\star \approx \frac{1 - e^{-\delta\tau/2}}{2 - e^{-\delta\tau/2}}
    \label{sstar}
\end{equation}
as stated in the main text.

\section*{Simulation details}
For all simulations in the main text, we have used the following parameters unless otherwise stated:
$\mu_L = 1/N, K(t=0)_i \!= \!1 \forall\,\, i$ and a fully connected network with $N=100$, $\delta=0.05$. We have found that adding small heterogeneity in each household's labor does not change the dynamics significantly and the equilibrium dynamics also remain the same for different initial capital distributions with different $\sum_i K_i = K$.
For Fig.\,1, savings rate distribution plots are shown at the final state of the simulation at time $5\tau \cdot 10^3$, far beyond the point where the model has reached its asymptotic dynamics. For each value of $\tau$, $200$ independent simulations are run and all values of $s_i$ are recorded for each $\tau$ to construct a histogram, which is normalized so  the values add to one for each set of simulations with the same $\tau$.
In Fig.\,4, Erdős–Rényi random graphs are constructed such that every node is connected to every other node to avoid households that are static and depreciation rate is $\delta =0.2$ to speed-up computation.

\section*{Savings rate dependency and topology}
\begin{figure}[h]
      \centering
        \includegraphics[width=0.35\textwidth]
        {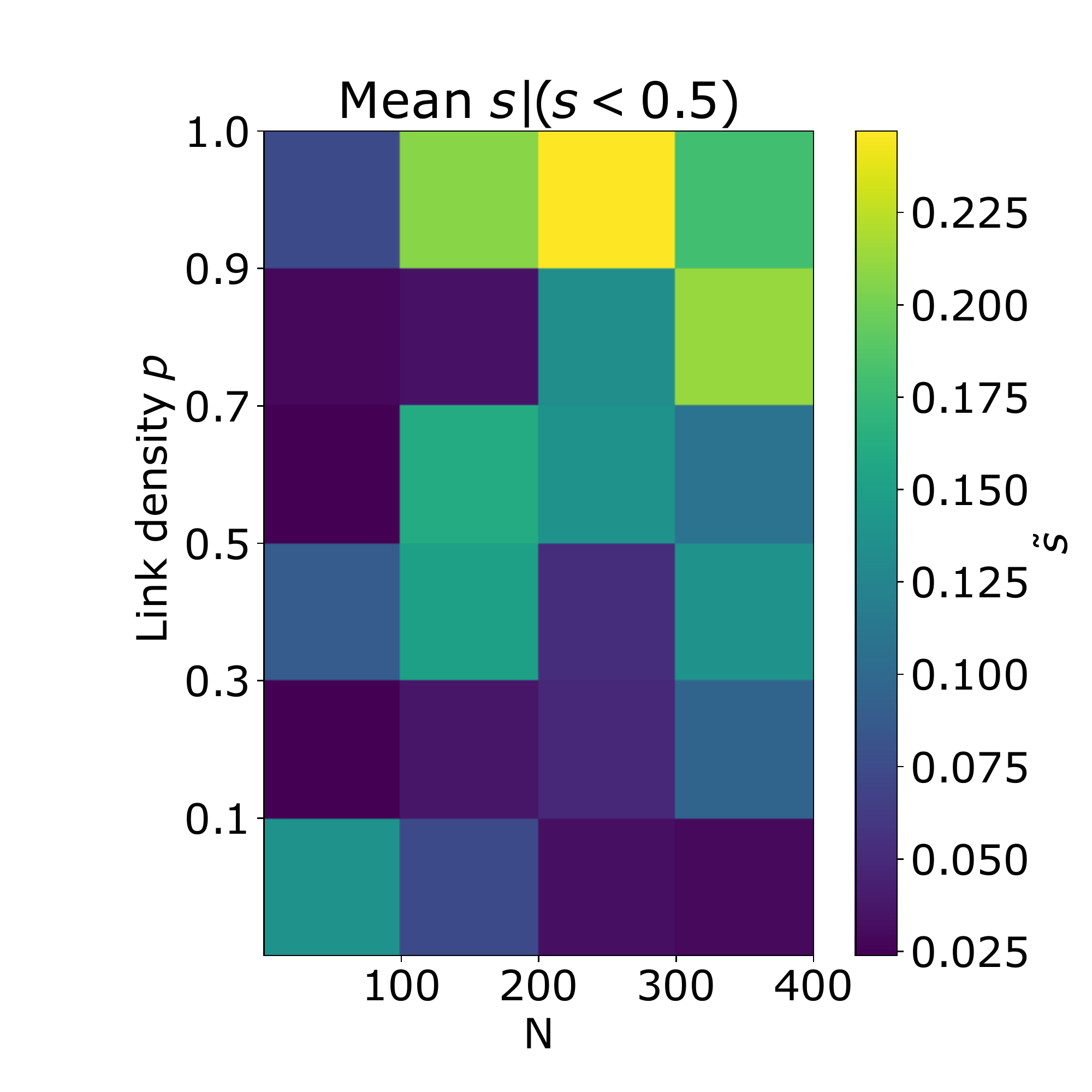}
        \includegraphics[width=0.35\textwidth]
        {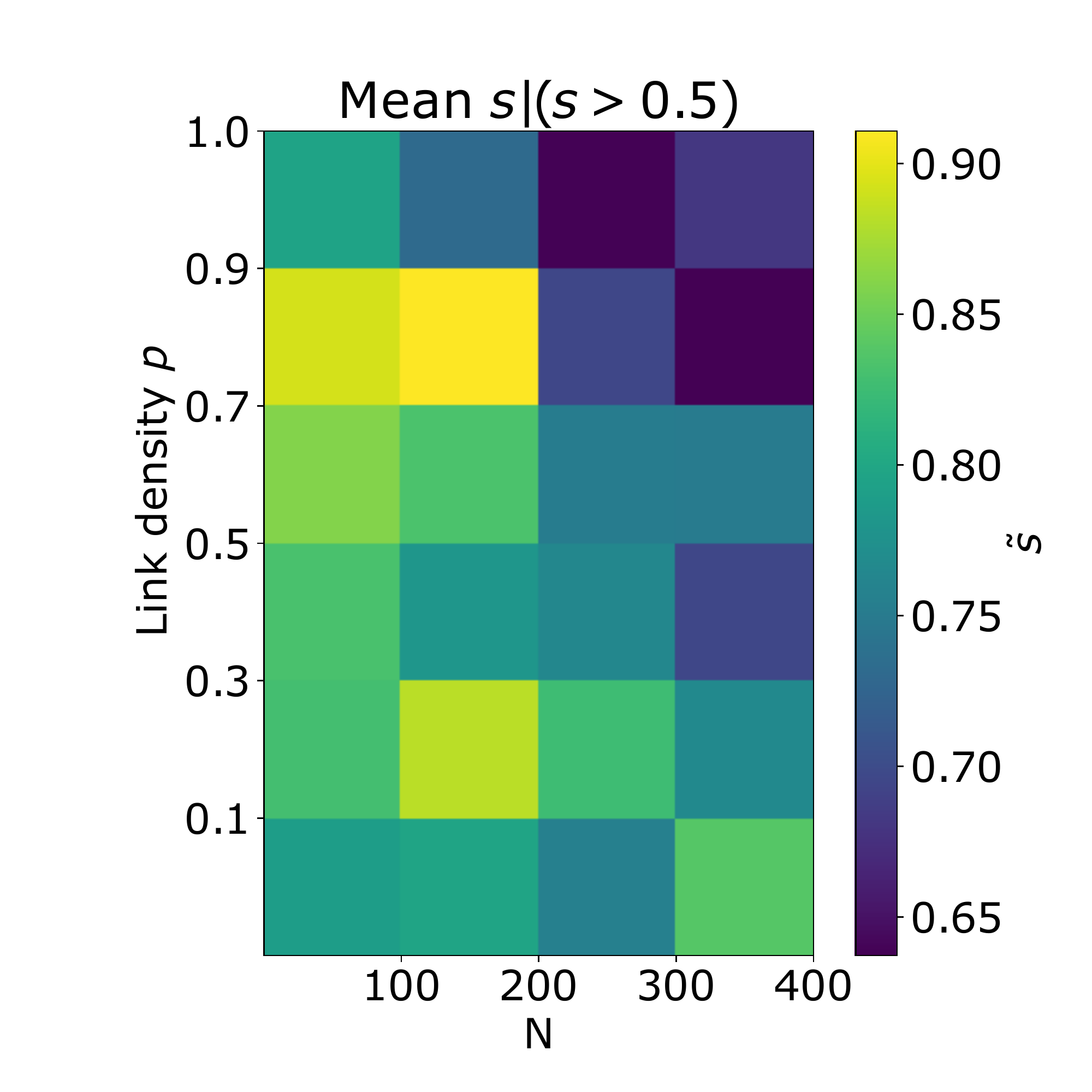}
	\caption{Dependence of the high and low savings rates on the network topology. We show the mean values of the savings rates in the lower mode with $s<0.5$, (left) and the higher mode with $s>0.5$ as a heatmap. We can see a range of different values that depend on the topology of the network.}
    \label{fig:savrate-topology}
\end{figure}

In Fig.\,\ref{fig:savrate-topology}, we show the mean values for aggregate savings rates for the two modes of the  oscillatory regimes.
We can observe values ranging from 2\%--23\% for the lower savings rates and from 65\% to 90\% for the higher ones.
Furthermore, we can see that the values for the high savings rates are strongly anticorrelated with the values of the lower savings rates, as they still average to $s^\star=0.5$.

\section*{Effective time-preference rate}
\begin{figure}[h]
      \centering
        \includegraphics[width=0.7\textwidth]
        {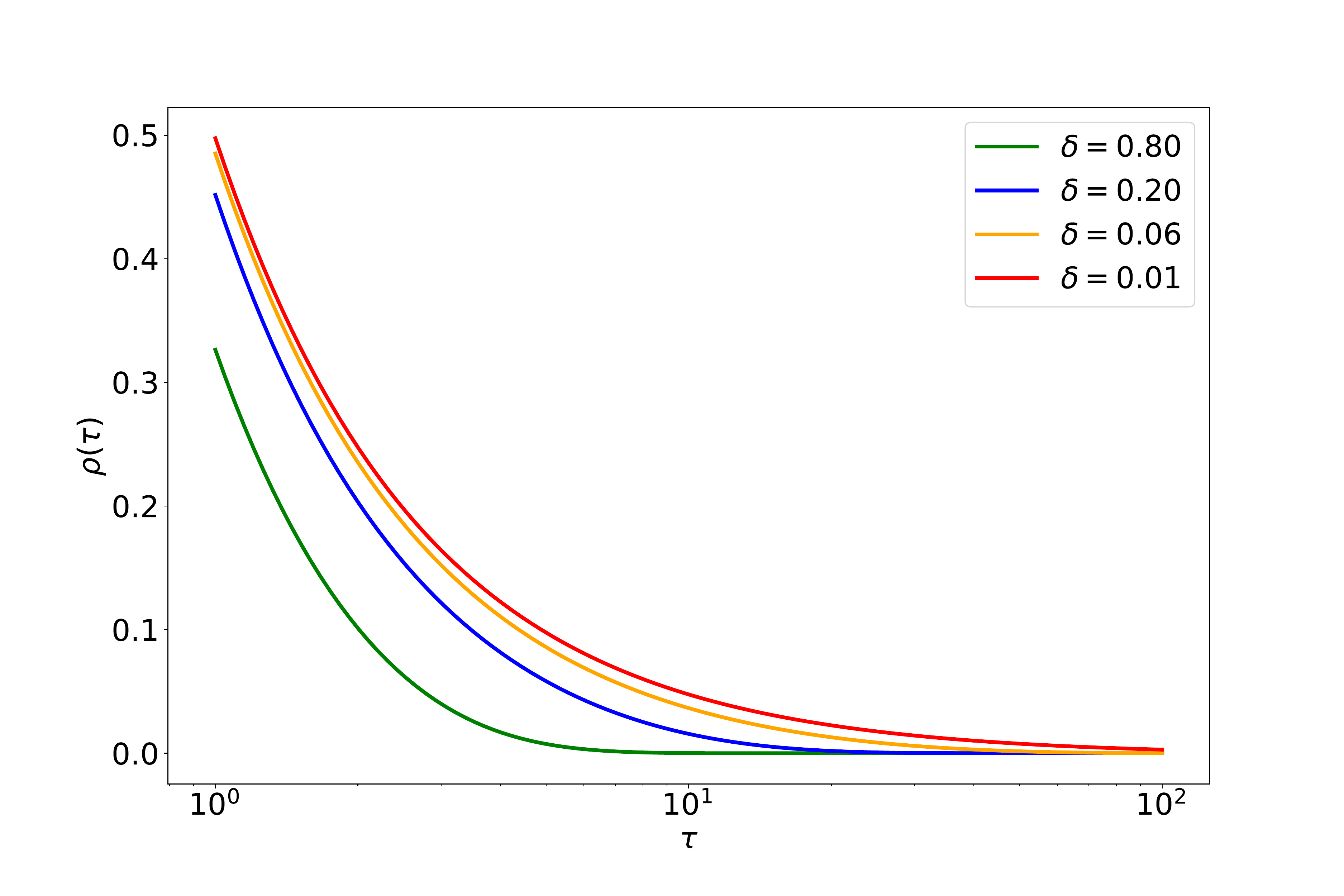}
	\caption[Effective discount rate as a function of the social interaction time]{Effective discount rate as a function of the social interaction time. The corresponding discount rate $\rho(\tau)= \frac{\delta}{2} ({e^{\delta \tau/2} -1})^{-1}$ of our model is shown for various depreciation rates $\delta$ and low social interaction times $\tau<\tau_\mathrm{c}$.}
    \label{rhotaufig}
\end{figure}

In the RCK model, the discount rate is a free parameter, and the representative household solves the intertemporal optimization problem by choosing a savings rate of
\[
s^\ast_\mathrm{RCK} = \frac{\alpha \delta}{\rho + \delta}
\]
(see Methods section from the main paper). 
In contrast, in the stable regime of our agent-based model, the myopic behaviour of the many individual households that learn their individual savings rates by social learning with a mean social interaction time $\tau$ leads to an aggregate savings rate $s$ that converges to a steady state $s^\star$ which is approximately
\[
s^\star_\mathrm{ABM} \approx \frac{1 - e^{-\delta\tau/2}}{2 - e^{-\delta\tau/2}},
\]
as derived above.
So, the main household characteristic that determines $s^\star_\mathrm{RCK}$ is the discount rate $\rho$, 
while the main household characteristic that determines $s^\star_\mathrm{ABM}$ is the social interaction time $\tau$.
Comparing the above two equations, we see that a given social interaction time $\tau$ in the ABM leads to approximately the same aggregate savings rate as the discount rate
\[
\rho(\tau)= \frac{\delta/2}{e^{\delta \tau/2} -1}
\]
in the RCK model. Conversely, a given discount rate $\rho$ in the RCK model leads to approximately the same aggregate savings rate as the social interaction time
\[
\tau(\rho) = \frac{2}{\delta}\ln\left(1 + \frac{\delta}{2\rho}\right)
\]
in the ABM.
This relationship between $\rho$ and $\tau$ is shown in Fig.\ \ref{rhotaufig} for various depreciation rates $\delta$. Higher depreciation rates lead to lower discount rates, which is due to the dependence of $s^\ast_\mathrm{ABM}$ on $\delta$ and the inverse relationship between the classical economically optimal savings rate $s^\ast_\mathrm{RCK}$ with $\rho$. 
Thus, a higher depreciation leads to an increasing preference for the future. While maybe counterintuitive at first, it can be explained by the fact that $s^\ast_\mathrm{RCK}$ is obtained from the optimization of a representative agent which requires a higher savings rate for higher rates of depreciation for optimal consumption.

In the case where either $\delta$ or $\tau$ are small or $\rho$ is large, 
we even have $\rho\approx 1/\tau$,
which means that the discount rate of the RCK model almost exactly corresponds to the social interaction {\em rate} of our model.

\section*{Single household free-riding\label{freerider_si_k0}}

We have seen above that, given an ensemble of households with an  aggregate savings rate $s$, the individual consumptions $C_i$ will approximately evolve as in Eq.\,\ref{eq:Citau}. 
If the ensemble of households is large enough, it will contain some household $j$ that uses a savings rate close to the value of $s_i$ which maximizes Eq.\,\ref{eq:Citau}.
Let us call the latter the {\em best response} to $s$ and denote it by $s_i^\star(s)$.
Since household $j$ is then the most-consuming member of the ensemble, the social learning implies that other households will start copying $s_j \approx s_i^\star(s)$, after which $s$ will move slightly towards $s_i^\star(s)$ as well. 
We have seen that when $s$ equals the value $s^\star$, also $s_i^\star = s^\star$, so that $s^\star$ is an equilibrium.
It turns out that for values of $s$ slightly above $s^\star$, we get $s_i^\star < s^\star$, and for values of $s$ slightly below $s^\star$, we get $s_i^\star > s^\star$. 
Since updates occur asynchronously, after some stochastic perturbation of $s$ away from $s^\star$, this implies that $s$ moves back towards $s^\star$ with some inertia, so that those households which imitate $s_i^\star$ first in such a situation have a temporarily larger consumption than the others.

\begin{figure}[h]
      \centering
        \includegraphics[width=0.7\textwidth]
        {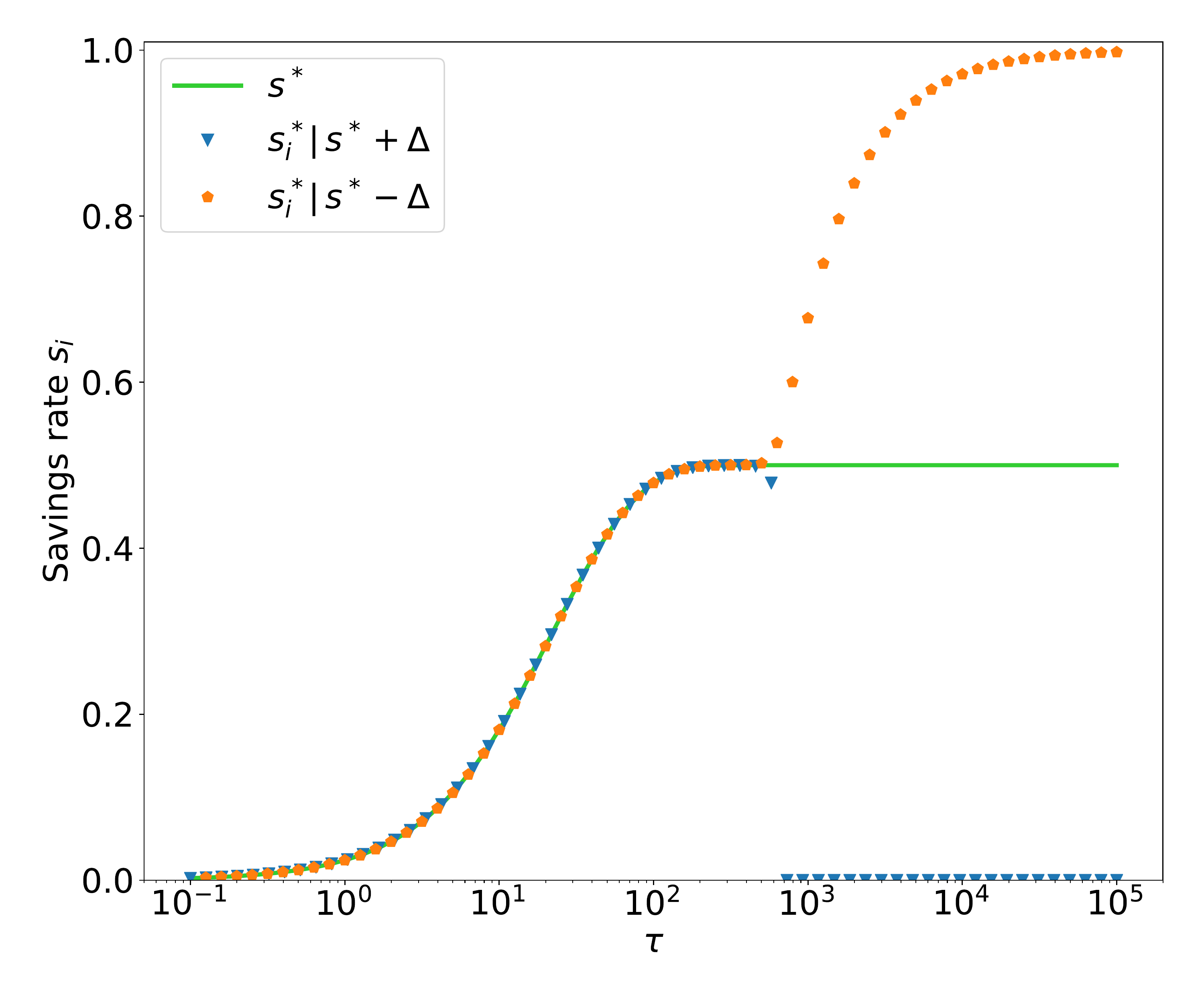}
	\caption{Best response dynamics. Optimal individual savings rate $s_i^\star$ (circles and triangles) given a very small perturbation of either $+\Delta$ (blue circles) or $-\Delta$ (orange triangles) of the aggregate savings rate $s$ away from its equilibrium value $s^\star$ (green line) as a function of the social interaction time $\tau$, for $\Delta = 10^{-7}$. 
    Below a critical value of $\tau$, best responses are very close to the equilibrium value, while above the critical value, they diverge quickly towards very small or very large values in the opposite direction of the perturbation, thereby stabilising $s$ but leading to temporary profits due to the countercyclical exploitation of large rents (if $s < s^\star$) or large wages (if $s > s^\star$).}
    \label{freerider_pm}
\end{figure}

Interestingly, the difference between the best response $s_i^\star(s)$ and the equilibrium $s^\star$ depends strongly and highly nonlinearly on $\tau$, as can be seen in Fig.~\ref{freerider_pm}. 
For $\tau$ below some critical value, $s_i^\star(s)$ is very close to $s^\star$, which explains why not only the aggregate savings rate $s$ but also all individual savings rates $s_i$ remain close to $s^\star$ in the stable regime of our model.
However, for $\tau$ above the critical value, $s_i^\star(s)$ is close to zero if $s > s^\star$
and very large if $s < s^\star$, converging to unity as $\tau$ increases. So as long as $s > s^\star$, more and more households will switch to $s_i\approx 0$, making $s$ decrease until $s < s^\star$, after which more and more households will switch to very large $s_i$, making $s$ increase again, leaving fewer and fewer households at intermediate values of $s_i$. This explains the emergence of two separate classes and extreme inequality in the oscillatory regime of our model.
In both cases, those households with a countercyclical savings rate can be seen as ``free-riding'' on the behaviour of the others. When $s < s^\star$, those households with large $s_i$ accumulate capital and get high rents due to an overall scarcity of capital. When $s > s^\star$, wages and thus incomes increase due to the overall abundancy of capital, so those household with small $s_i$ profit from consuming a larger share of their temporarily large income.

\section*{Dependence of the critical interaction time \label{meandegree}}
\begin{figure}[h]
\centering
        \includegraphics[width=0.7\textwidth]
        {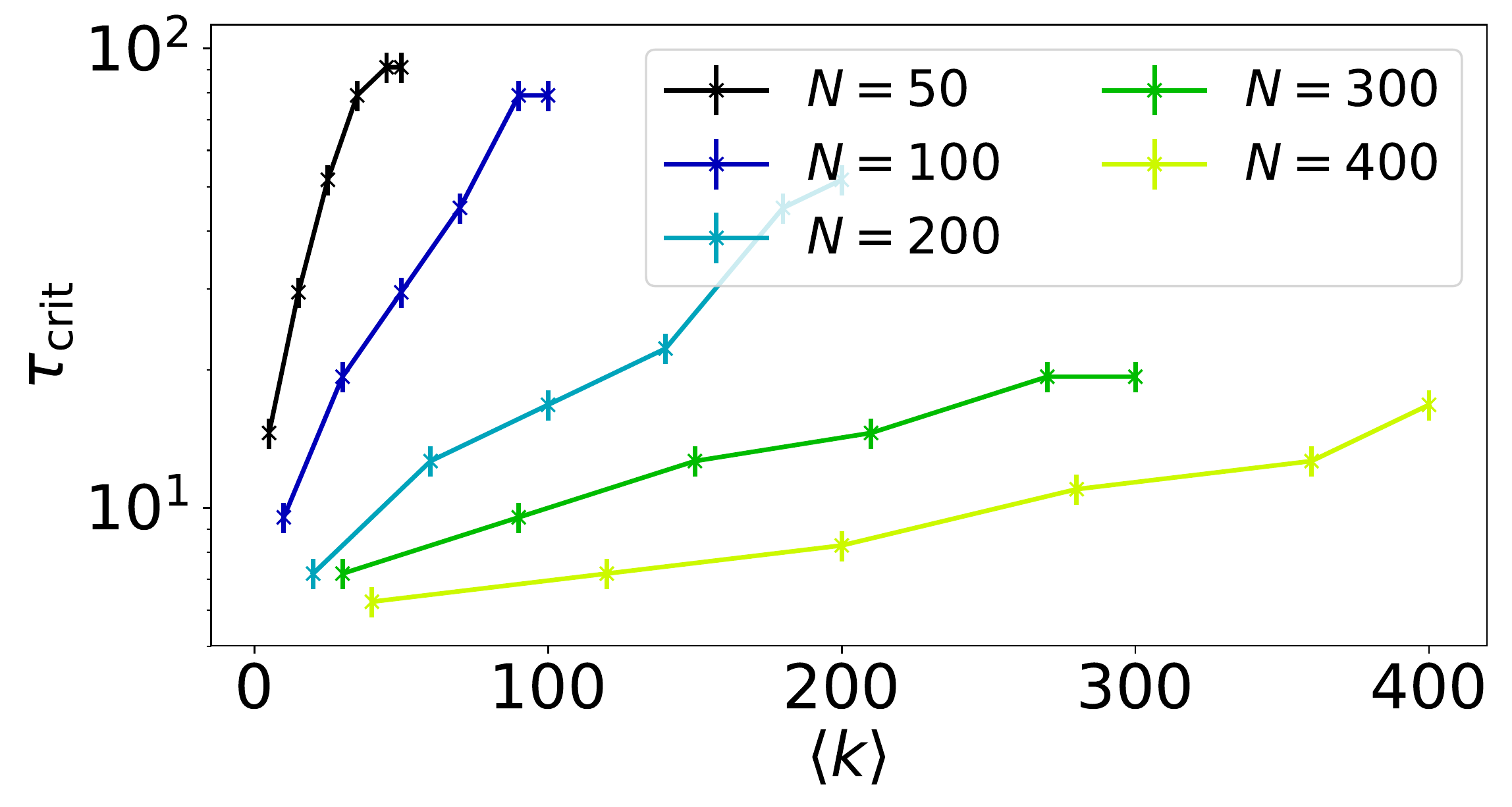}
	\caption{Critical interaction time $\tau_\mathrm{c}$ as a function of the network mean degree.  Here, $\tau_\mathrm{c}$ is shown as a function of network size $N$ for different mean degrees, given by $\langle k \rangle =Np$ for Erdős–Rényi graphs \cite{Erdos1959} and $\alpha=0.5, \delta=0.2$ and the critical rate increases with the mean degree $\langle k \rangle$. 
    \label{taucrit_2}}
\end{figure}

\begin{figure}[h]
\centering
        \includegraphics[width=0.6\textwidth]
        {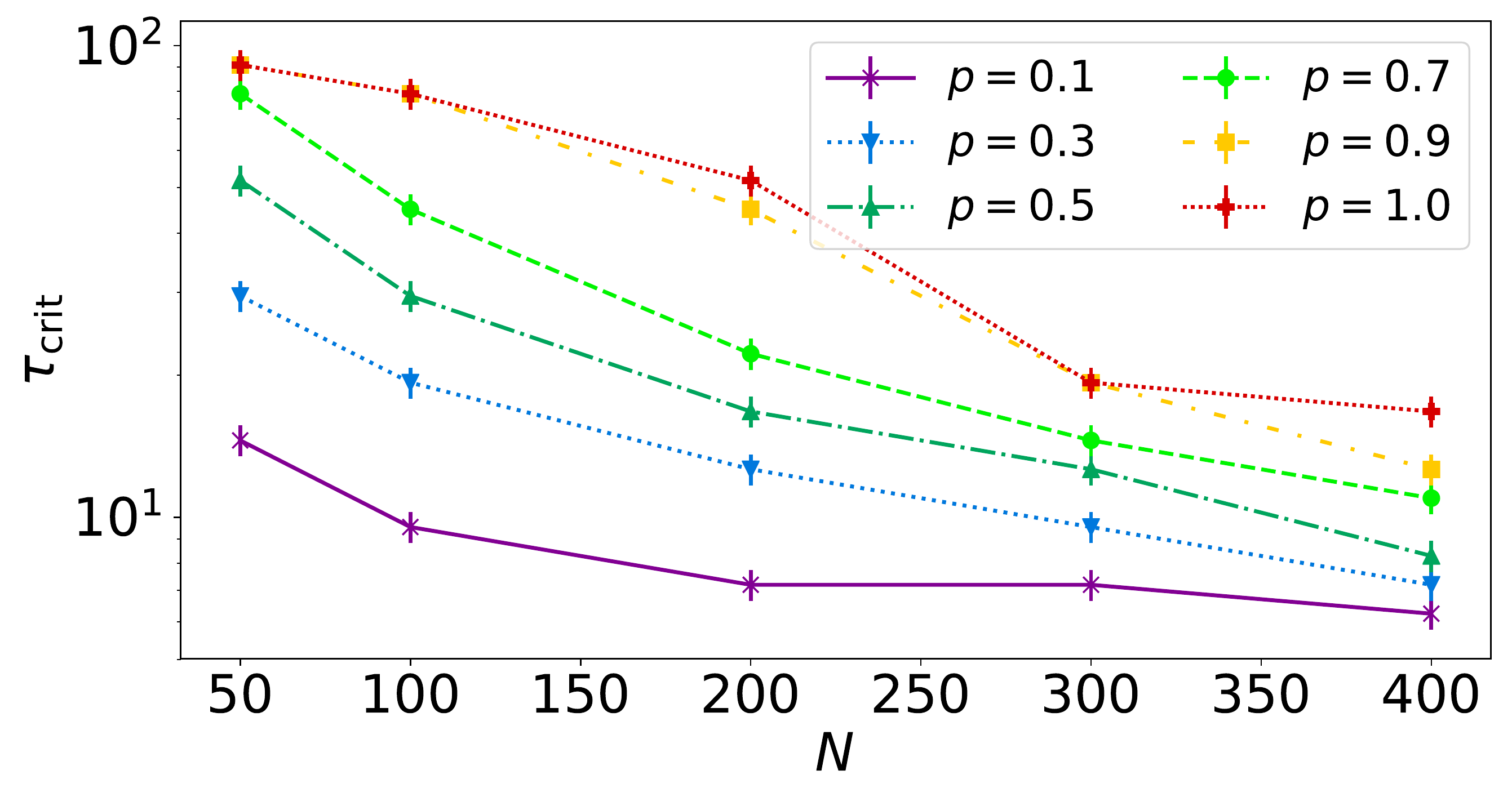}
	\caption{Critical interaction time $\tau_\mathrm{c}$ against number of households.  Here, $\tau_\mathrm{c}$ is shown as a function of network size $N$ for different link densities $p$ for Erdős–Rényi graphs \cite{Erdos1959} and $\alpha=0.5, \delta=0.2$. The critical interaction time scales proportionally with the number of households and inversely with the link density.
    \label{taucrit_3}}
\end{figure}

In Fig.~\ref{taucrit_2} and Fig.~\ref{taucrit_3}, we show an alternate views of Fig.~4 in the main text. Here we show the dependency of $\tau_\mathrm{c}$ on the mean number of neighbors $ \langle k \rangle$ and the number of households $N$ and the link density $p$. We see no clear alignment without using the average shortest path length as done in the main text.

\section*{Examples of qualitatively different oscillations}
\begin{figure}[h]
      \centering
        \includegraphics[width=0.7\textwidth]
        {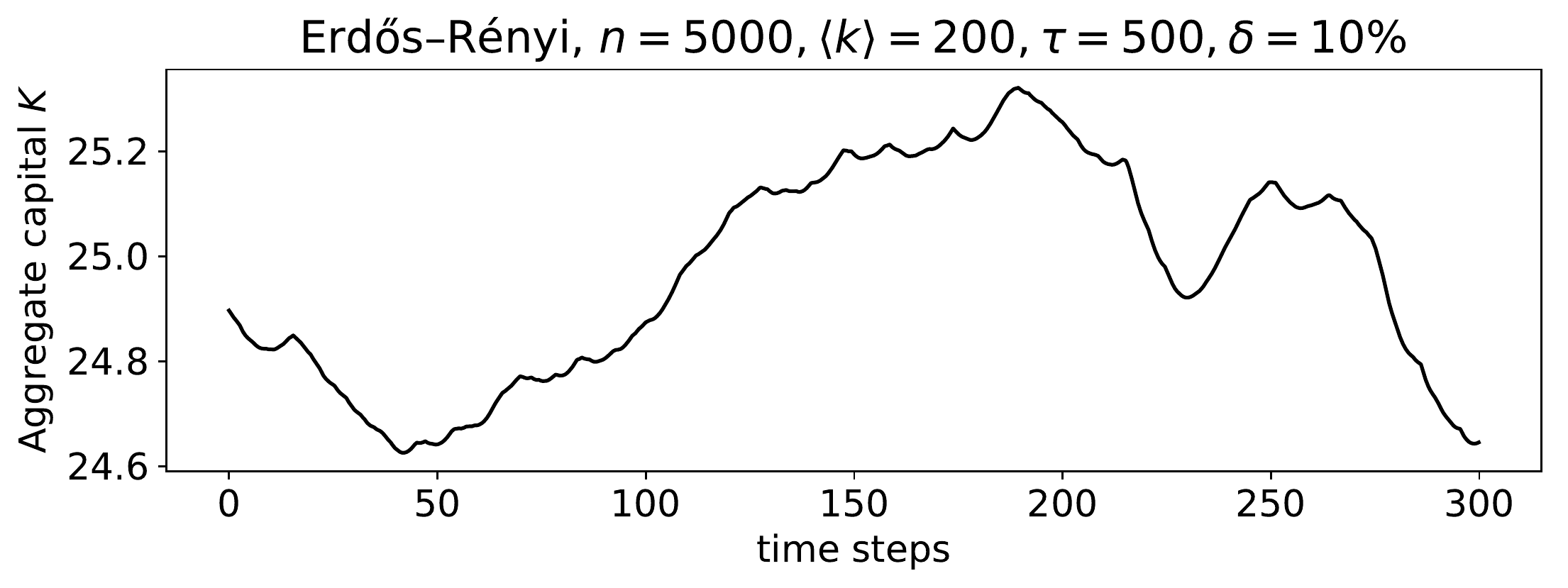} \\
        \includegraphics[width=0.7\textwidth]
        {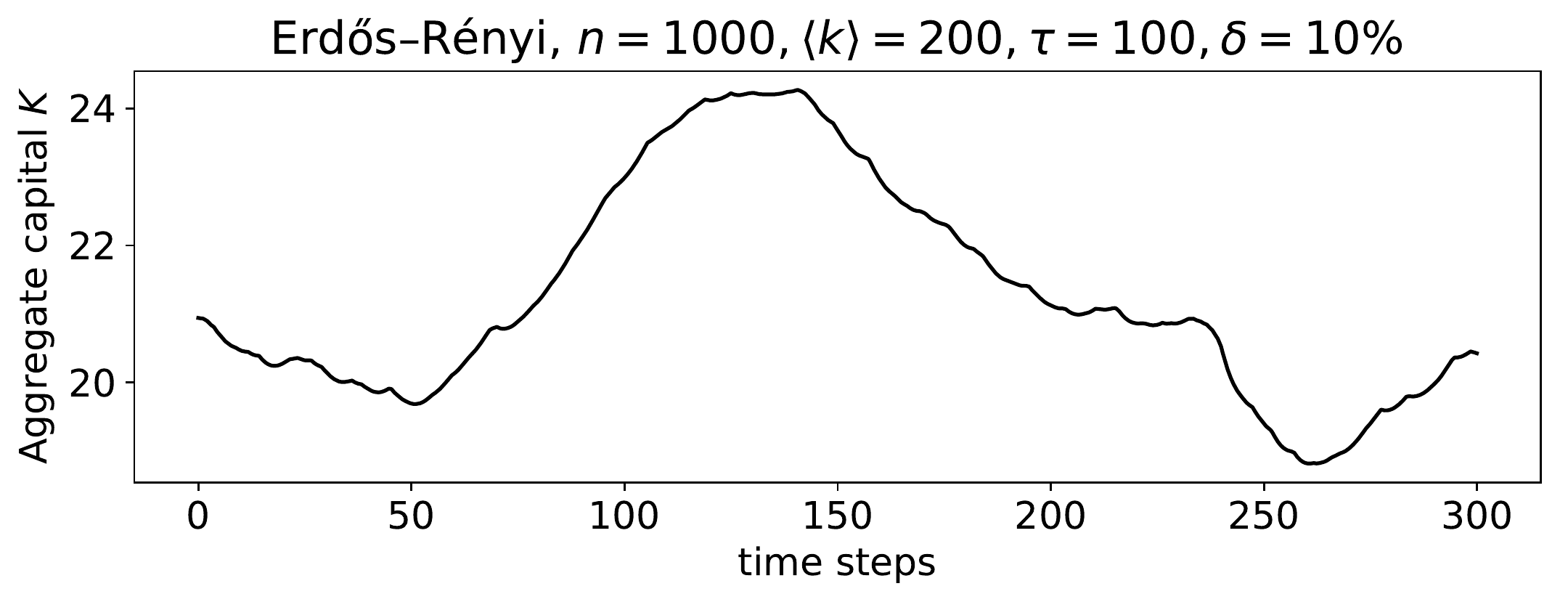} \\
        \includegraphics[width=0.7\textwidth]
        {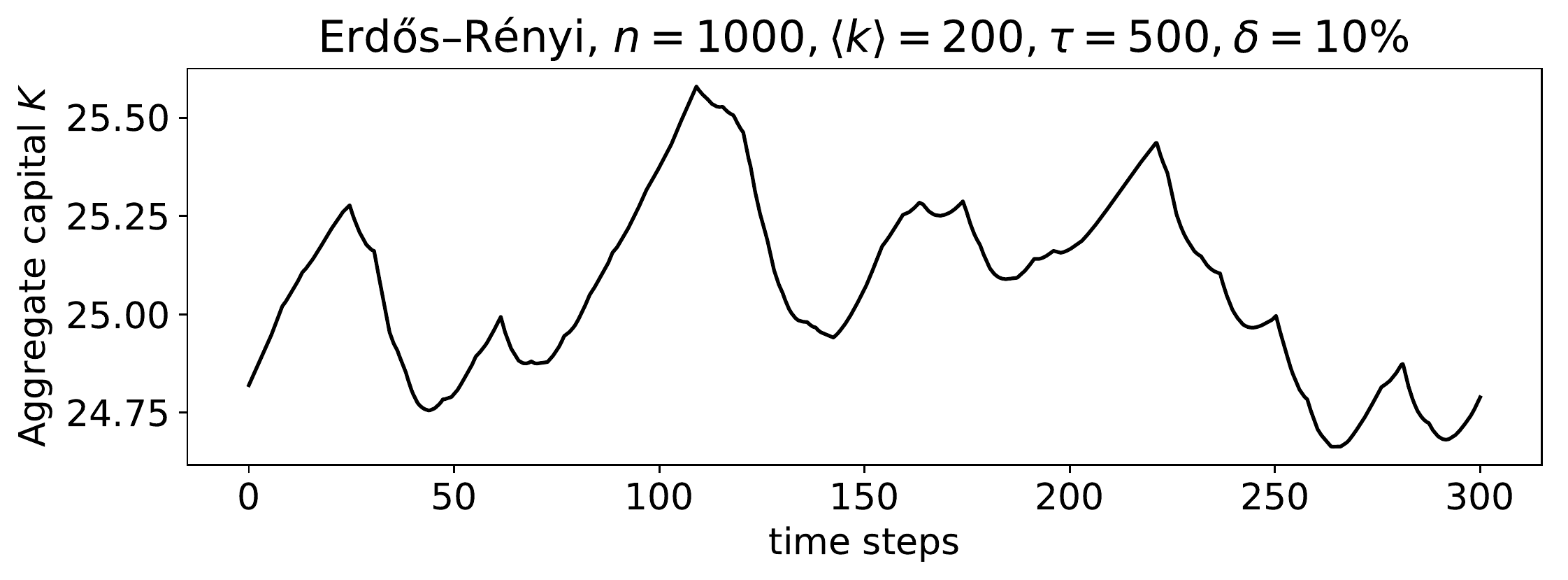} \\
	\caption{Various oscillation patterns for different Erdős–Rényi graphs. The parameters are given in the titles of the figures. We can see that decreasing the number of agents and decreasing the social interaction time $\tau$ increases the roughness of the aggregate capital curves. Note that labor is initialized as $1/N$ so that the equilibrium capital is constant for all simulations.}
    \label{fig:various-ER}
\end{figure}

\begin{figure}[h]
      \centering
        \includegraphics[width=0.7\textwidth]
        {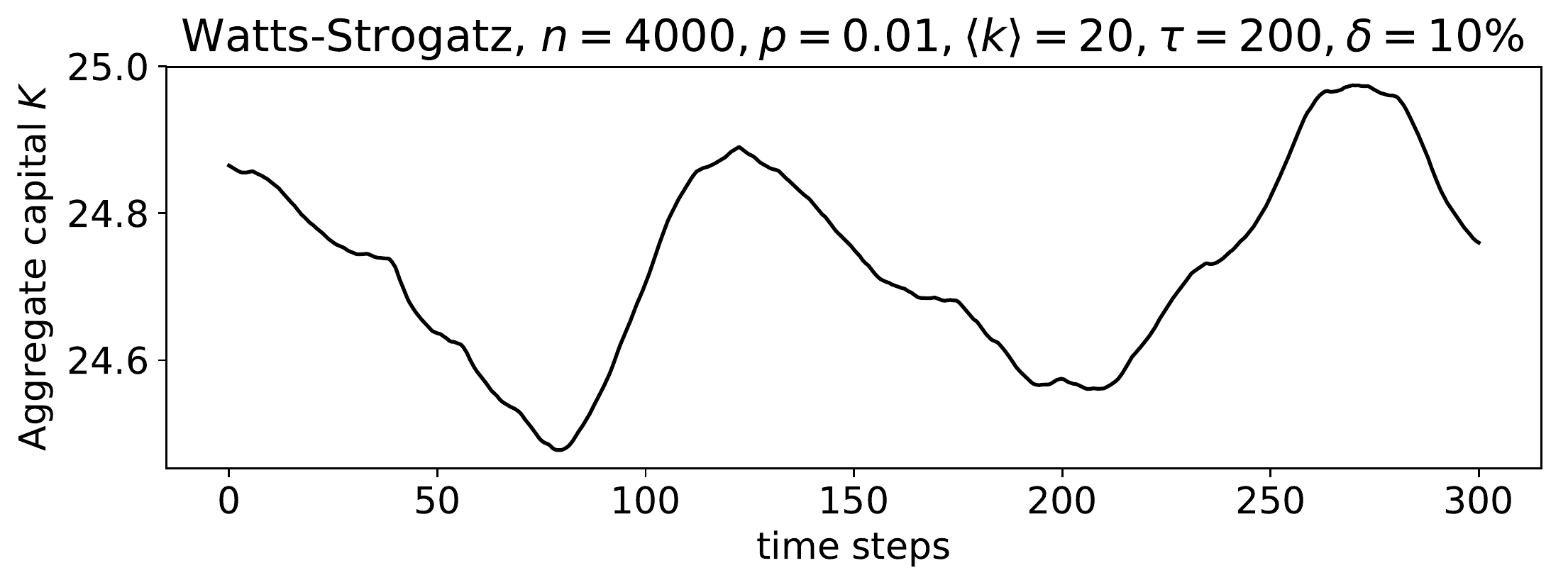} \\
        \includegraphics[width=0.7\textwidth]
        {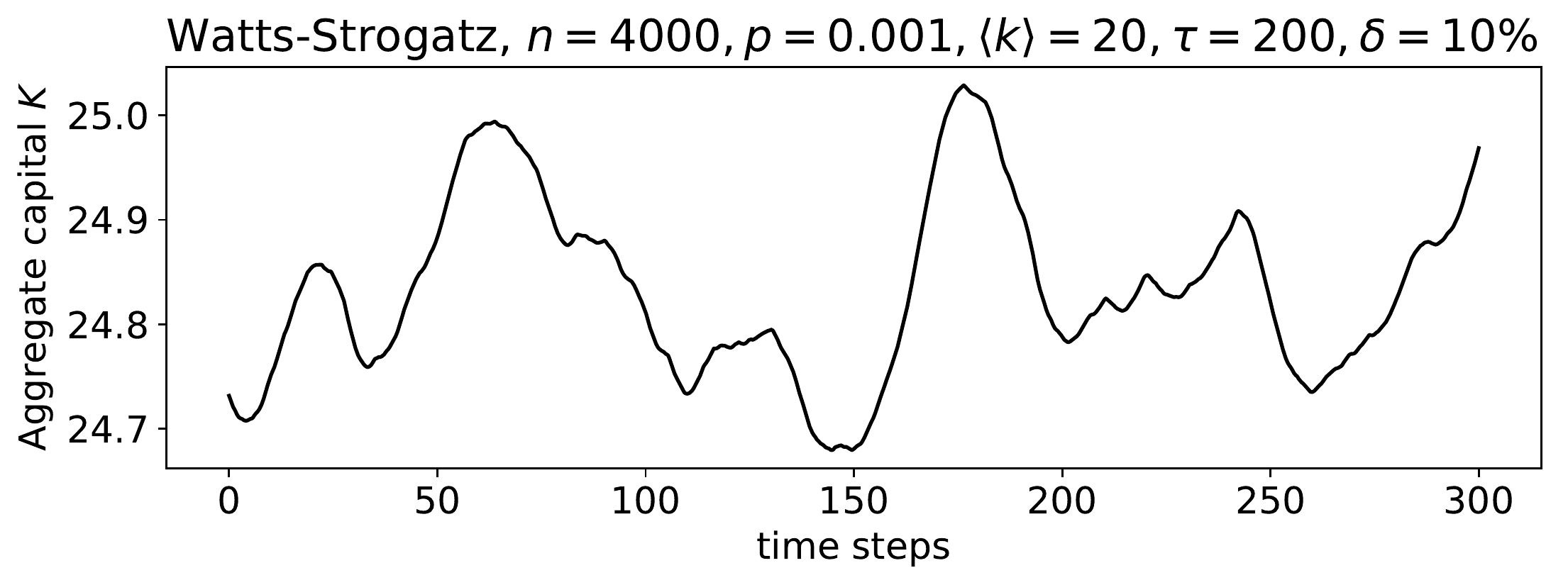} \\
	\caption{Various oscillation patterns for different Watts-Strogatz graphs \cite{watts1998collective}. The parameters are given in the titles of the figures. 
	We can see that decreasing the shortest paths length (by decreasing the rewiring parameter $p$, leads to a more rough aggregate capital curves with multiple frequencies.
	Note that labor is initialized as $1/N$ so that the equilibrium capital is constant for all simulations.}
    \label{fig:various-WS}
\end{figure}

Varying the network architecture and the number of agents simulated not only changes the value of $\tau_\mathrm{c}$ but also the frequency of the oscillations. 
We show  300 time step intervals in the oscillatory regime for  various simulations using  Erdős–Rényi random graphs \cite{Erdos1959} in Fig.~\ref{fig:various-ER} and Watts-Strogatz ``small world'' \cite{watts1998collective} networks in Fig~\ref{fig:various-WS}. 
We can find that decreasing the number of agents and the social interaction time $\tau$ increases the roughness of the aggregate capital curves indicating multiple oscillation frequencies with different amplitudes.

Using the depreciation rate per time step $\delta$ to roughly correspond to the real-word depreciation per year, we find that the large oscillation frequencies are generally in the range of 50-100 years and smaller oscillations (e.g. lower panel in Fig.~\ref{fig:various-WS}) in the range of 15-20 years. These results show that, while the main oscillation frequency is too low for a typical business cycle,  this model can generate various oscillations with different frequencies that are superimposed on each other, similar to economic waves.

\newpage
\textbf{Supplementary Figures}


\end{document}